\documentclass[runningheads]{llncs}
\usepackage{graphicx}
\usepackage{cite}
\usepackage{amsmath,amssymb,amsfonts}
\usepackage{algorithmicx}
\usepackage{textcomp}
\usepackage{xcolor}
\usepackage[hyphens]{url}

\usepackage{multirow}
\usepackage{float}
\usepackage{listings}

\usepackage{siunitx}
\usepackage{booktabs}
\usepackage{hyperref}

\usepackage{array}

\usepackage{comment}

\usepackage{siunitx}
\usepackage{fp}
\usepackage{wrapfig}
\usepackage{subcaption}
\usepackage{sidecap}
\usepackage{graphicx}

\newcommand{\numtomillion}[1]{%
  \FPdiv{\result}{#1}{1000000}%
  \num[round-mode=places, round-precision=1, scientific-notation=false]{\result} M%
}

\newcommand{\numtothousand}[1]{%
  \FPdiv{\result}{#1}{1000}%
  \num[round-mode=places, round-precision=0, scientific-notation=false]{\result} K%
}

\newcommand{\numtothousandNoK}[1]{%
  \FPdiv{\result}{#1}{1000}%
  \num[round-mode=places, round-precision=0, scientific-notation=false]{\result}%
}

\newcommand{\tabitem}{~~\llap{\textbullet}~~}

\usepackage{tikz}

\NewDocumentCommand{\langkeyword}{m}{%
    \texttt{\textcolor{violet!80!black}{#1}}%
}

\NewDocumentCommand{\langoperator}{m}{%
    \texttt{\textcolor{orange!60!black}{#1}}%
}

\newcommand{\PPAMsubmissionnumber}{}

\begin{document}
%
\title{Exploring the Design Space for Message-Driven Systems for Dynamic Graph Processing using CCA}
\titlerunning{Message-Driven Systems for Dynamic Graph Processing using CCA}
%
\author{Bibrak Qamar Chandio 
\and
Maciej Brodowicz
\and
Thomas Sterling}

%
\authorrunning{B. Chandio et al.}
%
\institute{Department of Intelligent Systems Engineering \\
 Indiana University Bloomington, USA \\
\email{\{bchandio,mbrodowi,tron\}@iu.edu}\\
}

\maketitle              
\begin{abstract}
Computer systems that have been successfully deployed for dense regular workloads fall short of achieving scalability and efficiency when applied to irregular and dynamic graph applications. Conventional computing systems rely heavily on static, regular, numeric intensive computations while High Performance Computing systems executing parallel graph applications exhibit little locality, spatial or temporal, and are fine-grained and memory intensive. With the strong interest in AI which depend on these very different use cases combined with the end of Moore’s Law at nanoscale, dramatic alternatives in architecture and underlying execution models are required. This paper identifies an innovative non-von Neumann architecture, Continuum Computer Architecture (CCA), that redefines the nature of computing structures to yield powerful innovations in computational methods to deliver a new generation of highly parallel hardware architecture. CCA reflects a genus of highly parallel architectures that while varying in specific quantities (e.g., memory blocks), share a multiple of attributes not found in typical von Neumann machines. Among these are memory-centric components, message-driven asynchronous flow control, and lightweight out-of-order execution across a global name space. Together these innovative non-von Neumann architectural properties guided by a new original execution model will deliver the new future path for extending beyond the von Neumann model. This paper documents a series of interrelated experiments that together establish future directions for next generation non-von Neumann architectures, especially for graph processing. 

\end{abstract}

\lstdefinelanguage{Racket}{
    morekeywords={struct, begin, predicate, propagate, diffuse, define, work, if, cond, let, let*, for-each, set!, cons, Integer, Float, Vector, Pointer, null},
    keywordstyle={\color{violet!80!black}},
    sensitive=true,
    morecomment=[l]{;},
    morestring=[b]",
    alsoletter={<, >, !,-},
    morekeywords=[2]{SSSP-Action, SSSP-Diffuse, edge-address, edge-weight, vertex-edges, vertex-level, set-vertex-level!, list, inform-neighbors, bfs-action, BFSDiffuse,  set-future!, allocate, insert-edge-action, insert-edge, vertex-has-room, vertex-ghost, vertex-score, set-vertex-score!, future-pending, enqueue-future!, future-pending!, enqueue!, rhizome-collapse, op, bcast, vertex-iteration-score, vertex-msg-count, set-vertex-msg-count!, inform-score-to-neighbors, vertex-out-degree, vertex-in-degree, germinate-action},
    keywordstyle=[2]{\color{orange!60!black}},
    morekeywords=[3]{mutable, rhizome-shared},
    keywordstyle=[3]{\color{blue}},
    literate={lambda}{{{\color{violet!80!black}$\lambda$}}}1 
              {call/cc}{{{\color{violet!80!black}call/cc}}}6 
              {eq?}{{{\color{orange!60!black}eq?}}}3 
              {?}{{{\color{orange!60!black}?}}}1 
              {'}{{{\color{red!75!black}\small\textbf{\textquotesingle}}}}1 
              {==}{{{\color{orange!60!black}==}}}2 
              {+}{{{\color{orange!60!black}+}}}1 
              {/}{{{\color{orange!60!black}/}}}1 
              {>}{{{\color{orange!60!black}>}}}1 
              {null?}{{{\color{orange!60!black}null?}}}5 
              {\#}{{{\color{blue}\#}}}1 
              {:}{{{\color{blue}:}}}1,
}

\lstset{
    language=Racket,
    basicstyle=\ttfamily\scriptsize, 
    stringstyle=\color{red}\ttfamily,
    commentstyle=\color{green!40!black}\ttfamily,
    morecomment=[l][\color{magenta}]{\#},
    showstringspaces=false,
    breaklines=true,
    numbers=left,
    numberstyle=\tiny\color{gray},
    numbersep=2pt,
    upquote=true,
}

\section{Introduction}
Graphs are emerging as the challenging form of computation for AI and similar problem types, not well served by conventional von Neumann architecture. With the end of Moore’s Law and Dennard scaling as well as the poor memory access patterns currently achievable, a new architecture has become essential for immediate growth in processing performance. This paper addresses this challenge and defines a new class of non-von Neumann architecture that replaces conventional practices for future ultra-high scale computing. Although of broader and more general purpose, the narrower domain of dynamic graph problems is the focus of this paper as these are of significant interest in the rapidly expanding field of AI and poorly match the processing characteristics of today’s computing systems.

Graphs appear throughout computing in a number of roles and can serve as the foundation for determining much of a computer’s operational properties. Here, five principal classes of graphs are introduced in Table \ref{tab:graphmani} although their relative degree of use varies significantly and directly impacts the roles and degree that they display. These are 1) data graphs, 2) data structures, 3) program graphs, 4) execution graphs, and 5) architecture graphs. In Table \ref{tab:graphmani}, the nature of each form and purpose of the types of graphs are described.  This paper introduces each kind although the architecture graph in the form of NoC, the data graph, and the means of employing the execution graph are the most significant for the purpose of this paper. Key to their interrelationship is that the data graph is dynamic and the execution graph (also dynamic) determines the active control, order, and concurrency of operations that are performed.  Although not shown here, a major contributing component is the runtime system discussed in a previous work  \cite{chandio2024rhizomes} along with details of the data structure that we use in this paper.

\setlength{\intextsep}{10pt} 
\begin{table}
  \caption{Graph Manifestations}
  \label{tab:graphmani}
  \centering

  \begin{tabular}{p{0.6in} p{4.2in}}
    \toprule
   
    \multicolumn{2}{c}{\textbf{Graph Type and Properties}}\\
    \midrule
   
    \multirow{2}{*}{\includegraphics[width=0.6in,height=0.6in]{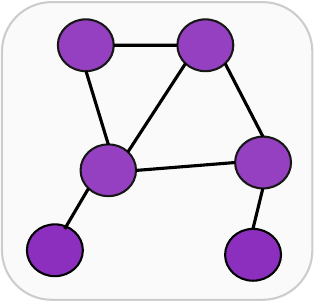}} & \textbf{Data graph} \\
    & \tabitem It contains the input data.\\
    & \tabitem Parallelism intrinsic in topology.\\
    & \tabitem Static or dynamic.\\
    & \tabitem Regular or irregular (variable degrees).\\
    \midrule

    \multirow{2}{*}{\includegraphics[width=0.6in,height=0.6in]{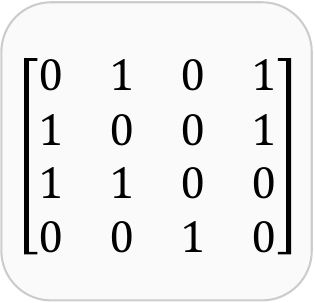}} & \textbf{Data-structure of the Data graph} (Adjacency Matrix shown)\\
    & \tabitem Adjacency List or Adjacency Matrix.\\
    & \tabitem Vertex objects using pointers. A single vertex can be partitioned hierarchically in a tree and operated upon using recursive parallelism.\\
    \midrule

    \multirow{2}{*}{\includegraphics[width=0.6in,height=0.6in]{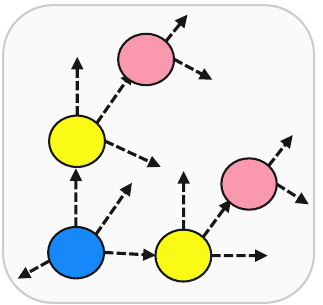}} & \textbf{Program graph} (BFS shown)\\
    & \tabitem Dataflow graph.\\
    & \tabitem Programmer describing the algorithm.\\
    & \tabitem Compiler further changes/adapts it.\\
    \midrule

    \multirow{2}{*}{\includegraphics[width=0.6in,height=0.6in]{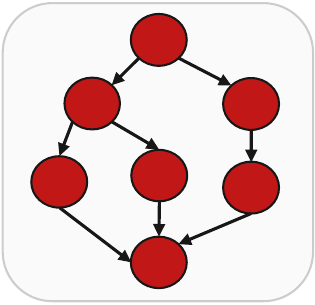}} & \textbf{Execution graph} (DAG shown)\\
    & \tabitem Variable with respect to each instance.\\
    & \tabitem Dynamic due to runtime control.\\ 
    & \tabitem Time dependent.\\
    \midrule
    
    \multirow{2}{*}{\includegraphics[width=0.6in,height=0.6in]{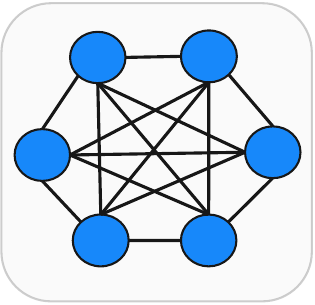}} & \textbf{Architecture graph} (Full Mesh shown)\\
    & \tabitem Network-on-Chip.\\
    & \tabitem System area network topology.\\
    & \tabitem Routing Algorithms.\\
    
  \bottomrule
\end{tabular}
\end{table}

Physical memory in the Continuum Computer Architecture (CCA) \cite{CCA2018} is logically unified across the entire scalable system. Naming (address space) is essentially accessible from any action in the system, independent of relative physical placement. But significant, sometimes dramatic, variation in distances travelled through the system between the source and destination locations of the global memory cannot be mitigated by conventional methods including (but not limited to) multi-layer caches. CCA employs asynchronous actions instead a significant departure from usual practices.  Any two data elements both launched simultaneously may arrive at their destination at very different times because of their highly varied distances. The arrival times between two interacting data elements can be both nearby but differ substantially in up stream launch times. Again, CCA treats these global data similarly but deals with such discrepancies on the fly.  This is handled by the usage of futures mechanisms managed by hardware within the architecture.

CCA communication is unified by creating a global and shared network that is built for a large number of small messages; exactly the opposite of usual methods. This is a departure from staging data at various levels of hierarchies to optimize for bandwidth. Previous methods enforced different, and mostly static and synchronous control flow mechanisms. To export high degrees of parallelism and hide latency, communication must be made transparent and captured through the asynchronous and dynamic parallel semantics that can enable spawning computations from within data object. This principle is used for the mode of programming. It spawns dynamic computations by moving “work to data” in the form of active messages, called \textit{actions}, for fine-grain compute and communication parallelism.

The CCA execution model allows global parallelism. Most contemporary computing architectures are based on von Neumann sequential machines that are connected together, thereby creating what is called the Communicating Sequential Processes (CSP) execution model. Such machines may have high degrees of local parallelism, but in practice, they employ globally sequential methods. Section \ref{subsec:theoretical} further discusses some important prior work in this regard. Section \ref{sec:hardware-design} inherits some of these ideas in designing the hardware structure of our system.

\begin{figure}
  \centering
  \begin{subfigure}{.49\linewidth}
    \centering
    \includegraphics[width=\linewidth]{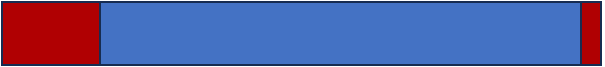}
    \caption{One fork-join with 20\% serial and 80\% parallel portions in the code.}
    \label{fig:parallelism-amdahl-simple}
  \end{subfigure}
  \hfill
  \begin{subfigure}{.49\linewidth}
    \centering
    \includegraphics[width=\linewidth]{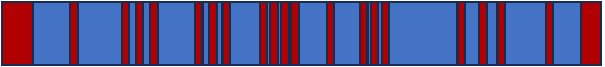}
    \caption{Many fork-joins with the same 20\% serial and 80\% parallel portions in the code.}
    \label{fig:parallelism-amdahl-fine}
  \end{subfigure}
  \caption{Execution signatures showing serial and parallel portions. Red is serial and blue is parallel.}
  \label{fig:parallelism-amdahl}
\end{figure}

Graphs do not have well-structured spatial locality patterns. Techniques such as data-parallelism do not inherently scale well. Data-parallelism techniques are generally static and regular, requiring data that is not only spatially contiguous but also remains unchanged once the computation begins. Graph operations are also not highly compute intensive. Consequently, control (or task) parallelism techniques, such as those under fork-join and Bulk Synchronous Parallel (BSP), incur significant synchronization overheads and execution irregularities. This is explained with computations containing execution signatures shown in Figure \ref{fig:parallelism-amdahl-simple}. Large bulk parallel portions are scaled by offsetting the amount of parallelism with coarser grain computations. Such methods, when employed with fine-grain computations with execution signatures of Figure \ref{fig:parallelism-amdahl-fine}, not only limit parallelism but also introduces side-effects, such as the straggler effect \cite{StanglerEffect}, where the entire computation waits for the slowest strand (process) to make progress. While coarse-grain computations have the margin to tolerate these effects, fine-grain computations are more sensitive. Departing from coarser-grain to fine-grain parallelism, for increase in orders of magnitude of parallelism and yet achieving scalability, will demand new approaches that will combine aspects of data parallelism and control parallelism.

\section{Prior Techniques and their Synthesis}\label{sec:prior-work}
This section discusses some of the past contributions, techniques, and ideas that we believe have the potential to be synthesized to enable future scalable, fine-grain, and dynamic computational models and architectures. 
\subsection{Theoretical Works}\label{subsec:theoretical}
Typically, contemporary computing systems are conceived under the sequential process computing model, where instructions are issued one at a time. Although local optimizations such as superscalar, pipelines, out-of-order execution, and branch prediction have been made to extract parallelism and are coupled with the CSP model for scaling, nevertheless, there still remains an inherent lack of global parallelism at scale.

Functional and concurrent models provide unbounded parallelism and can be synthesized, especially utilizing the properties of lambda calculus \cite{Backus1978}, actors model \cite{Actors2010}\cite{ActorsAgha1985} and cellular automata \cite{Wolfram1982CellularAA}. For example, currying in lambda calculus has the properties of:

\begin{enumerate}
\item Allowing the computation to progress while waiting on some arguments to be available.
\item Allowing the runtime context of the computation to be stored and potentially migrated to some other part of the larger physical computing machine.
\end{enumerate}

The above two properties are of interest to problems that arise out of the irregular, fine-grain and dynamic nature of graph processing. These can be used to: 1) increase throughput of the system by evaluating computations and not waiting on some dependencies, and 2) migrate computations to improve load balance of the system and also hide latency by migrating computation to where the data resides.

Properties of the actors model include being history-sensitive (unlike dataflow functional systems) and being able to gracefully deal with dynamic topology, thereby making it re-configurable and extensible. From both a structural and functional point of view, cellular automaton provides neighborhood state transforms, which can inspire future class of Processing In Memory (PIM) architectures. These can be used to further evaluate the placement, capacities and capability, and interaction of compute and memory logic. Section \ref{sec:hardware-design} contains details about these ideas and their incorporation into our system.

Linial's LOCAL model \cite{LinialLocal} studies graph algorithm that can provide an output based on the local vertex data and the data surrounding a smaller neighborhood in the graph topology. It models the computations as if the vertex has some processing ability and can send and receive messages from their neighbors. In this way the LOCAL model theoretically studies graph algorithms as if the graph were executing itself. In the language of computer architecture, this appears to be close to some form of PIM-styled message-driven computing system.

\subsection{Architecture and Machines}
High-Level Language Computer Architectures (HLLCA) \cite{HLLCA1998}, such as Intel iAPX \cite{IntelIapxGarbageCollection}, and knowledge-based systems like the NETL architecture \cite{NETL1983}\cite{NETL2006} provide native support for garbage collection and application representation (think of the \verb|list| in Lisp). HLLCAs also incorporate tagged memory \cite{Tagged1973}, which enables a single operation or instruction to handle many types, resulting in compact code representation \cite{lispbenchmark1985}. Tagged memory maybe more important for message-driven systems that send work to data by having smaller instruction code size for message movement. The prior work and ideas of HLLCAs can be used to unite the application, compiler, OS, and the architecture for runtime introspection and on the fly optimization, which is a unique demand for dynamic and irregular applications like dynamic graph processing.

Furthermore, PIM architectures not only provide an answer to the von Neumann bottleneck, but also together with capabilities such as execution migration, of active messages \cite{ActiveMessages1992}, parcels and thread percolation \cite{HTMA1999}, can explore more fine-grain parallelism and hide latency. These along with the Dataflow architectures’ properties of exposing fine-grain parallelism, through the program control graph \cite{ArvindDataflow1986}, and J-Machine's \cite{Jmachine1993} built-in support for fine-grain asynchronous messaging and a system model for actors, can provide a way towards uniting data parallelism with control parallelism. 

Tesseract \cite{Tesseract2015} uses 3D stacked technology the Hybrid Memory Cube (HMC) to further the promise of PIM. Graph data can reside on any cube and can be referenced or mutated by the means of function invocation. These function calls can be blocking or non-blocking. In the latter case, the programmer must ensure global barrier synchronization to guarantee whether a non-blocking call finished. This implies that Tesseract implements the BSP model and can be challenging to implement globally asynchronous regimes. 

Dalorex \cite{Dalorex2023} is a cache-less architecture, where new tasks spawn at each memory indirection. These tasks are then sent for execution to the cores where the data resides. Dalorex's programming model and execution can support fully asynchronous computations without any global barriers. It uses CSR format with loop based programming constructs that make it less flexible for dynamic graph computations.

\subsection{Networking Infrastructure and Algorithms}
Fine-grain computations impose different kind of challenges on achieving a scalable networking infrastructure. These include the ability to:
\begin{enumerate}
\item Allow small size messages to be sent with low latency; as sending messages in bulk or blocking communication can reduce the amount of parallelism exposed.
\item Allow large number of small messages; the network must be able to sufficiently deal with resources waiting for contention.
\item Route a large number of small messages in a smaller number of hops.
\end{enumerate}

Networks such as the Data Vortex, and low diameter topologies such as Kautz, and PolarFly \cite{polarfly2022} try to address these three challenges. Data Vortex provides a low latency and contention free network \cite{datavortexrandom2016} that is important for implementing message-driven fine-grain architectures at scale. The Kautz and PolarFly topologies provide the properties of small diameter needed to route messages in a small number of hops. The data graphs themselves have diameters and the idea is to provide the architecture graph the capability to physically traverse the diameter in as little hops as possible (comparing the logical data graph diameter). On the other hand, more pertinent to the scope of this paper, is the network-on-chip (NoC). For example, the most common NoC, the mesh, although high in bandwidth, suffers from high latency due to its larger diameter. This can be mitigated with techniques such as wrapping around the borders to form Torus-Mesh, or adding extra links to the mesh topology such as the Ruche networks \cite{Ruche2020}, there by lowering the diameter and effectively the latency.

\section{Hardware Design Space Exploration}\label{sec:hardware-design}
As our exploration of hardware design space is inspired by CCA, we organize the system as an interconnection of homogeneous global memory Compute Cells (CCs), which are capable of; 1) data storage; 2) data manipulation; and 3) data transmission to adjacent CCs. Since a single CC has a limited amount of memory, an arrangement of CCs tessellated, for tight coupling, and interconnected work together to provide logical unification, larger data storage capacity, and parallelism. Such an interconnect, in its simplest form, is a mesh network, and will be based on shape of the tessellation. The actual number of CCs, their capacity and capabilities, and the mesh arrangement is a design objective to optimize for.

\subsection{Design Parameters for a Message-Driven Chip}\label{subsec:design-cca}
\begin{figure}
  \begin{minipage}[c]{0.35\textwidth}
    \includegraphics[width=2.2in]{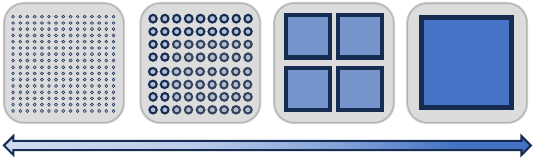}
  \end{minipage}\hfill
  \begin{minipage}[c]{0.53\textwidth}
    \captionsetup{skip=0pt}
    \caption{Idea of the compute continuum. Storage and computation happens in a medium of computing cells. The medium can be a continuum of a large number of tiny computing cells to a single big cell.}
    \label{fig:continuum}
  \end{minipage}
\end{figure}

The shape and capabilities of a single CC along with how these CCs are aggregated together to form a chip is a design compromise between three parameters: memory, compute, and communication. For a given area, in the theoretical limit, at the one end lies an infinite number of infinitesimally small CCs and at the other end there is only a single big CC encompassing the entire area. The later can be synonymous to a conventional big core processor. In this sense there is a continuum from the very small to the large as depicted in Figure \ref{fig:continuum}

\subsubsection{\textbf{Shape and Communication:}}\label{subsec:shape-comm}
\begin{figure}
  \centering
  \begin{subfigure}{0.32\linewidth}
    \centering
    \includegraphics[width=\linewidth]{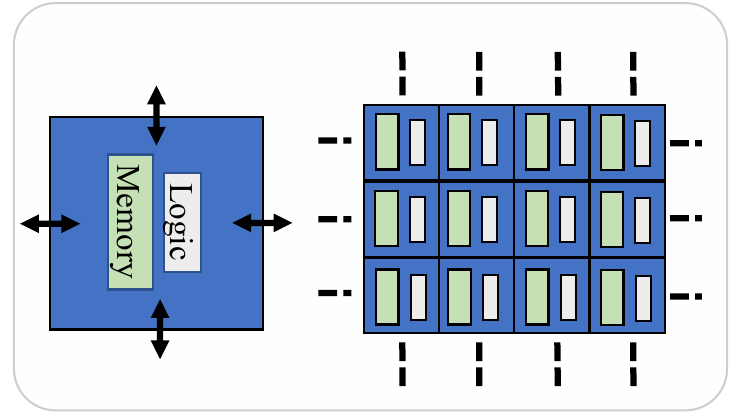}
    \caption{Square.}
    \label{fig:cc-square}
  \end{subfigure}
  \hfill
  \begin{subfigure}{0.32\linewidth}
    \centering
    \includegraphics[width=\linewidth]{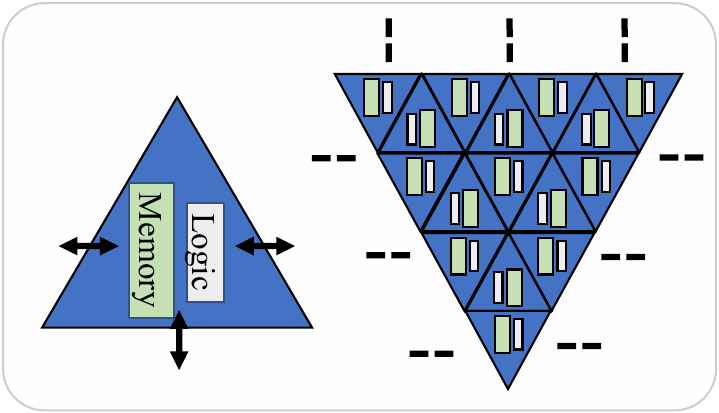}
    \caption{Triangular.}
    \label{fig:cc-triangle}
  \end{subfigure}
  \hfill
  \begin{subfigure}{0.32\linewidth}
    \centering
    \includegraphics[width=\linewidth]{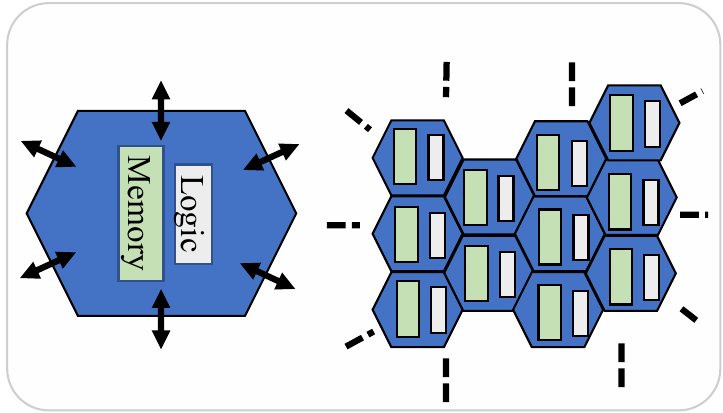}
    \caption{Hexagonal.}
    \label{fig:cc-hexagonal}
  \end{subfigure}
  \caption{Shapes of Compute Cells (CC) and their mesh tessellations.}
  \label{fig:cc_types}
\end{figure}

The shape of the CC determines its neighbor connectivity as shown in Figure \ref{fig:cc_types}. For example, a CC that is shaped as a square can form four communication channels to its neighbors when placed in a 2D tessellation. Other shapes can include: equilateral triangle with three channels, or hexagon with six channels. Shape of the CCs with more channels have higher communication bandwidth and message-delivery parallelism at the cost of more die area. The cost of more channels per CC is a compromise against the size of memory per CC, compute logic per CC, and the overall number of CCs in the entire chip. Section \ref{subsec:memory-compute} provides discussion on this trade-off space.

The shapes are not limited to regular tilings and 2D tessellations. There can be other complex non-regular tilings and 3D stacked tessellations; however, they are beyond the scope of this paper.

\subsubsection{\textbf{Memory and Compute:}}\label{subsec:memory-compute}

A single CC contains a limited amount of SRAM memory along with some computational ability. The capacity and capability of memory determines how much percent of die area is allocated to memory. Likewise, a CC has a limited number of computational resources such as arithmetic unit and other functional units that occupy the die space. Increasing these resources will either come at the expense of less memory capacity, and/or communication logic, and/or the total amount of CCs in a single chip.

\begin{figure}
  \begin{minipage}[c]{0.32\textwidth}
    \includegraphics[width=2in]{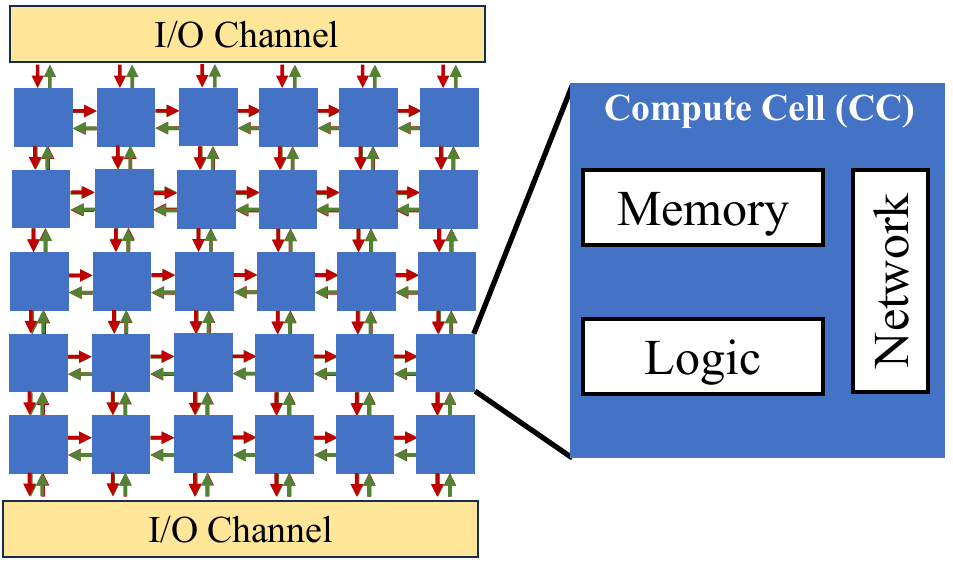}
  \end{minipage}\hfill
  \begin{minipage}[c]{0.55\textwidth}
    \captionsetup{skip=0pt}
    \caption{A $5\times6$ chip shown as an exemplar. Compute Cells containing local memory along with computing logic are tessellated in a mesh network.}
  \label{fig:cc}
  \end{minipage}
\end{figure}

To evaluate the practical impact of individual CC component sizes on the overall compute and communication capabilities of a chip of square CCs as shown in Figure \ref{fig:cc}, we estimated transistor counts related to primary functional units of the cell. The derivation of relevant geometric parameters assumed a TSMC N7 ($7$nm) process for chip implementation with the effective transistor density of $91$ million/mm$^2$. The central component of CC is a dual-port SRAM with 8T bitcells, structured as $4$ independently addressed banks with $64$-bit data I/O each and yielding a flit size of $256$ bits for message content transfer every clock cycle. While the memory I/O width remains constant in experiments, the address size is changed to reflect different memory capacities per CC along with corresponding transistor counts for address decoder, bit line drivers, sense amps, and output registers. The execution unit consists of instruction queue, instruction decoder, scalar ALU and FPU, and execution logic. All but the last require a fixed number of transistors of about $100K$ in aggregate; execution logic changes are due to different address register widths determined by the cell's memory size. The execution unit connects to the SRAM block via $64$-bit wide multiplexers that enable fetching instruction operands from any of the $4$ memory banks and permit delivery of ALU/FPU result to any of the banks. Finally, mesh links, their associated FIFO storage, flit multiplexers, and router logic are analyzed. The transistor count here varies due to FIFO depth and number of outgoing links. While other combinatorial logic, such as various state machines, handshake and arbitration support, etc. was not included in the totals, its impact is less than $1$--$2$\%.

\begin{figure}
  \centering
  \includegraphics[width=4.2in]{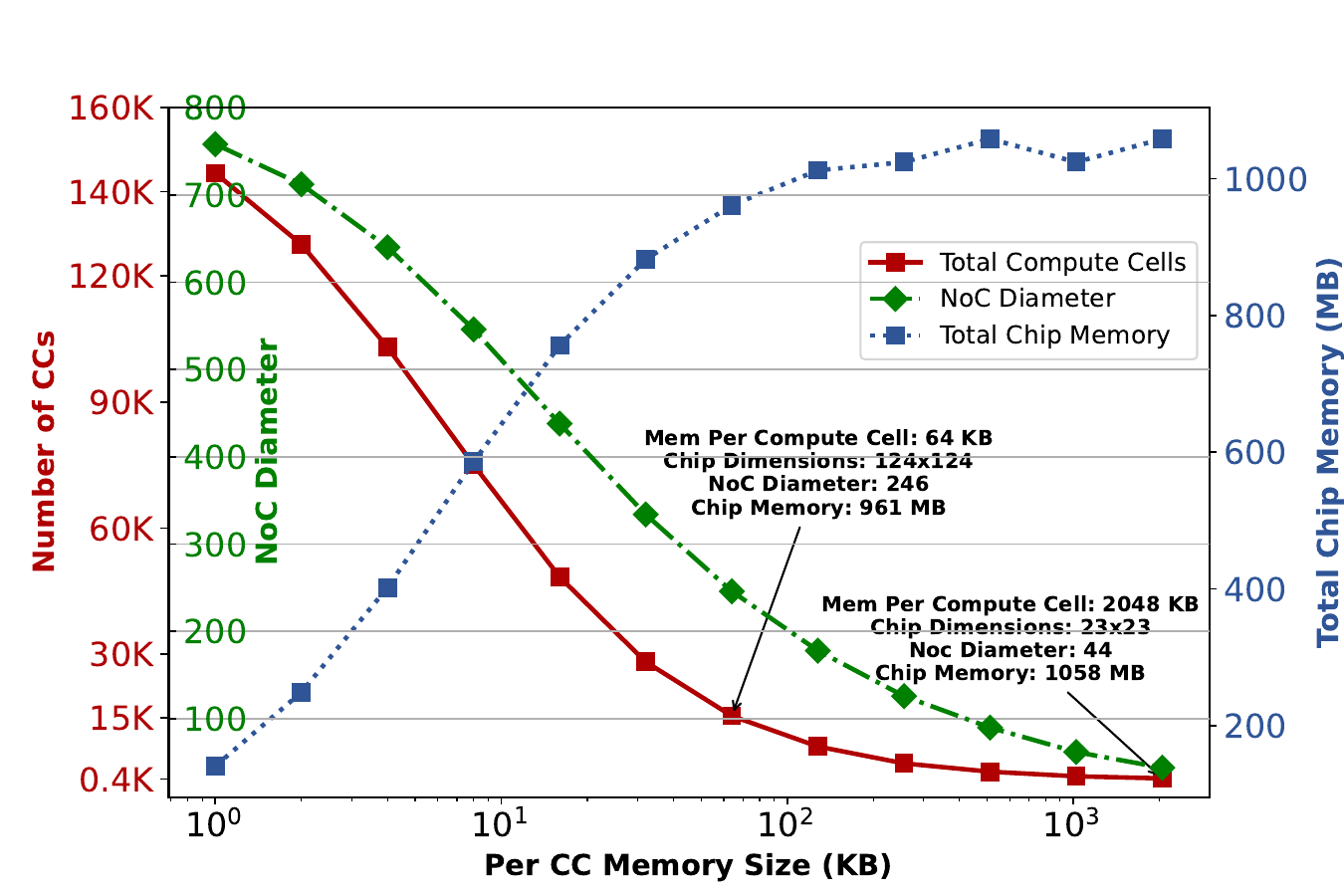}
  \caption{Relationship between number of mesh connected square shaped Compute Cells (CCs), mesh NoC diameter, and the total chip memory capacity in under a $306 mm^2$ area. Some data points are annotated with details so as to capture more context. X-axis in log scale.}
  \label{fig:chip-density-memory-relationship}
\end{figure}

Using the above estimates, Figure \ref{fig:chip-density-memory-relationship} shows the relationship between the total number of compute cells, the resultant mesh NoC diameter, and the total memory of a chip inscribed within an area of less than $306 mm^2$. It shows the trade-off space between the amount of compute parallelism, message-delivery latency, and the total chip memory capacity when per CC memory is increased or decreased.

\section{Batched Dynamic Graphs}

The system is natively designed to support dynamic graph processing using programming abstractions of \textit{actions}, and dynamic storage scheme of Recursively Parallel Vertex Object (RPVO) \cite{chandio2024rhizomes} that is constructed from vertex objects that are linked together hierarchy. These vertex objects contain a chunk of edges stored as pointers to other RPVOs representing other vertices of the graph. The root vertex object serves as the user program's accessible address for the vertex. As a vertex grows with the insertion of new edges, the RPVO can easily grow by adding new objects in its hierarchy. These objects may be allocated on the same CC or on a different CC. In the former case the hierarchical/recursive parallelism is serialized.

\begin{lstlisting}[language=Racket,caption={An action that implements the Breadth First Search. Some details are left out for brevity.},label=lst:cca-bfs-action,xleftmargin=0.3cm, escapeinside={*@}{@*}]
;; Breadth First Search Action
(define bfs-action ;; v: target vertex, lvl: level in
  (lambda ([v : (Pointer vertex)] [lvl : Integer])
    (if (> (vertex-level v) lvl)
      (begin
        ;; Perform work.
        (set-vertex-level! v lvl)
        ;; Diffusion occurs.
        (for-each 
          (lambda (e)
            ;; Get address of vertex along edge e.
            (let ([addr (edge-address e)])
              ;; Send action along edge e.
              (propagate bfs-action (list addr (+ lvl 1)))))
        (vertex-edges v))
      ))))
\end{lstlisting}

\textit{Actions} carrying edge(s) for insertion can be coupled with instructions to spawn new \textit{actions} thus enabling dynamic graph processing. We implement Dynamic-BFS using this approach as shown in Listing \ref{lst:cca-insert-edge}. When a new edge is inserted in a vertex it also germinates (inserts) a \langoperator{bfs-action}, of Listing \ref{lst:cca-bfs-action}, into the action queue of that CC. In subsequent cycles, the system executes this action and sends the BFS level along the newly added edge. In this way, Dynamic-BFS reuses prior results from prior computation without needing to compute levels for each and every vertex from scratch. 

\begin{lstlisting}[language=Racket,caption={Pseudocode for an action that inserts an edge in a vertex along with an application action, in this case the \langoperator{bfs-action}, on the compute cell containing the vertex.},label=lst:cca-insert-edge,xleftmargin=0.3cm, escapeinside={*@}{@*}]
(define insert-edge-action ;; v: target vertex, e: edge
  (lambda ([v : (Pointer vertex)] [e : edge])
    (begin
      (insert-edge v e) ;; Insert the edge in vertex v.
      ;; Germinate the `bfs-action` action in the compute cell where 
      ;; vertex v exists. It will be sent to the vertex point by edge e.
      (germinate-action bfs-action (list e (+ (vertex-level v) 1)))
    )))
\end{lstlisting}

\subsection{Experiments}
We implement our batched dynamic BFS using the CCASimulator \cite{ccasimulator:online} from \cite{chandio2024rhizomes}. In a single simulation cycle, a message carrying an \textit{action} can traverse one hop from one CC to a neighboring CC. Simultaneously, it can perform computing instructions, which are contained in the \textit{action} code or the creation of a new message when \langkeyword{propagate} is called. In this regard the \langoperator{bfs-action} can take $2$--$3$ cycles of compute, and when it creates and sends new \textit{actions} it takes cycles in the amount of edges contained in that vertex. Our batched dynamic BFS implementation is available at \cite{ccasimulator:online}.

\begin{table*}
  \caption{Details of the Input Dynamic Graphs}
  \label{tab:graphdetails}
  \centering  
  \begin{tabular}{|c|c|c|c|c|c|c|c|c|c|c|c|c|}
    \toprule
    \hline    
    \textbf{Sampling} & & \multicolumn{10}{|c|}{\textbf{Thousands of Edges Per Increment}} & \textbf{Final} \\
    \cline{3-12}
    \textbf{Type} & \textbf{\# V} & \textbf{1} & \textbf{2} & \textbf{3} & \textbf{4} & \textbf{5} & \textbf{6} & \textbf{7} & \textbf{8} & \textbf{9} & \textbf{10} & \textbf{Edges}\\
    \hline\hline
    Edge & \numtothousand{50000} & \numtothousandNoK{101682} & \numtothousandNoK{102012}  & \numtothousandNoK{101772} & \numtothousandNoK{101916} & \numtothousandNoK{101634} & \numtothousandNoK{101254} & \numtothousandNoK{101809}  & \numtothousandNoK{102076} & \numtothousandNoK{101645} & \numtothousandNoK{102239} & \numtomillion{1018039} \\
    \hline
    Snowball & \numtothousand{50000} & \numtothousandNoK{37315} & \numtothousandNoK{29238}  & \numtothousandNoK{47983} & \numtothousandNoK{68183} & \numtothousandNoK{87863} & \numtothousandNoK{108642} & \numtothousandNoK{129477}  & \numtothousandNoK{149413} & \numtothousandNoK{169416} & \numtothousandNoK{190509} & \numtomillion{1018039} \\
    \hline
    Edge & \numtothousand{500000} & \numtothousandNoK{1016373} & \numtothousandNoK{1016853}  & \numtothousandNoK{1015533} & \numtothousandNoK{1018007} & \numtothousandNoK{1018340} & \numtothousandNoK{1017923} & \numtothousandNoK{1016834}  & \numtothousandNoK{1019103} & \numtothousandNoK{1016846} & \numtothousandNoK{1018701} & \numtomillion{10174513} \\
    \hline
    Snowball & \numtothousand{500000} & \numtothousandNoK{222847} & \numtothousandNoK{328912}  & \numtothousandNoK{513890} & \numtothousandNoK{709723} & \numtothousandNoK{904420} & \numtothousandNoK{1101941} & \numtothousandNoK{1297078}  & \numtothousandNoK{1501559} & \numtothousandNoK{1698228} & \numtothousandNoK{1895915} & \numtomillion{10174513} \\
    \hline
    \bottomrule
    \multicolumn{13}{l}{\footnotesize There are ten increments to the graph each inserting a number of new edges.}\\
    \multicolumn{13}{l}{\footnotesize K is thousand, and M is million.}\\
  \end{tabular}
\end{table*}

\textbf{Datasets:} We perform our experiments using synthetic dynamic graphs from MIT's Streaming GraphChallenge \cite{StreamingChallenge2017}\cite{StreamingDataSets}. Table \ref{tab:graphdetails} provides details of the graph datasets used in our dynamic graph experiments. The graphs are constructed using two types of sampling methods: Edge and Snowball. In edge sampling, the edges are inserted as if they were formed or observed in the real world, while in Snowball sampling, the edges are inserted as they are discovered from a starting point \cite{graphChallenge2017}.

\begin{figure}
  \centering
  \begin{subfigure}{0.49\linewidth}
    \centering
    \includegraphics[width=\linewidth]{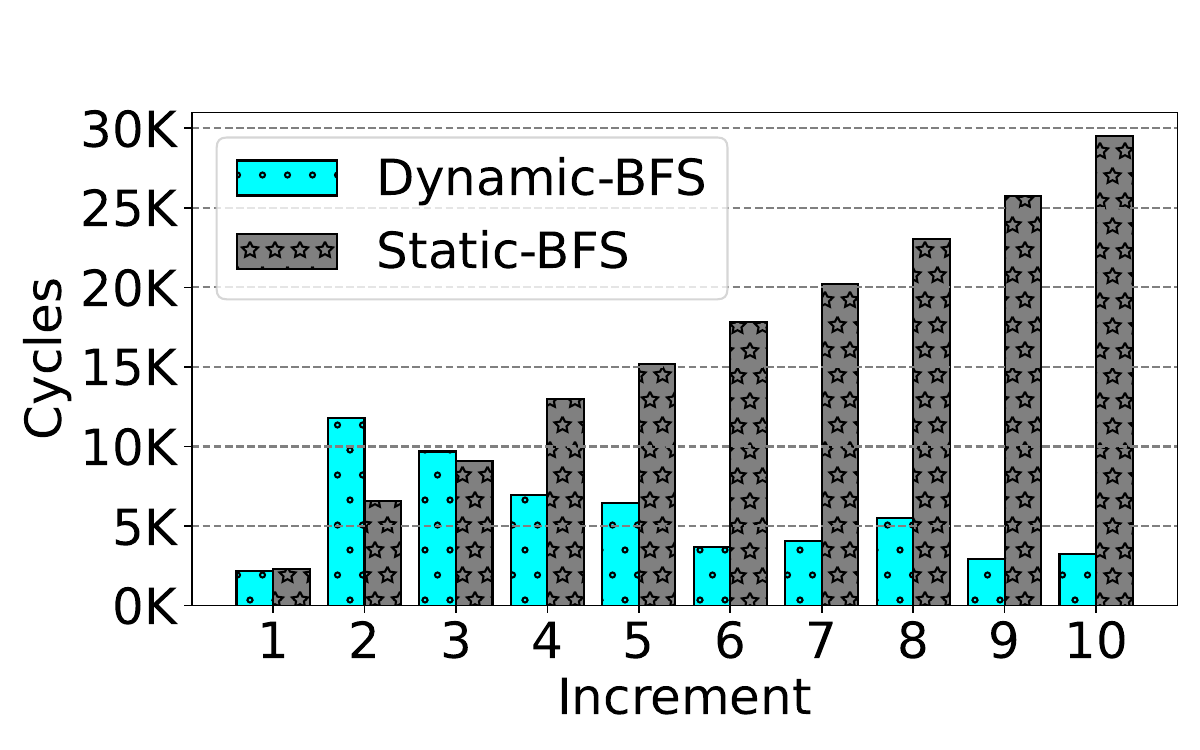}
    \caption{Edge Sampling \#V: 50K.}
    \label{fig:edge-sample-50k}
  \end{subfigure}
  \hfill
  \begin{subfigure}{0.49\linewidth}
    \centering
    \includegraphics[width=\linewidth]{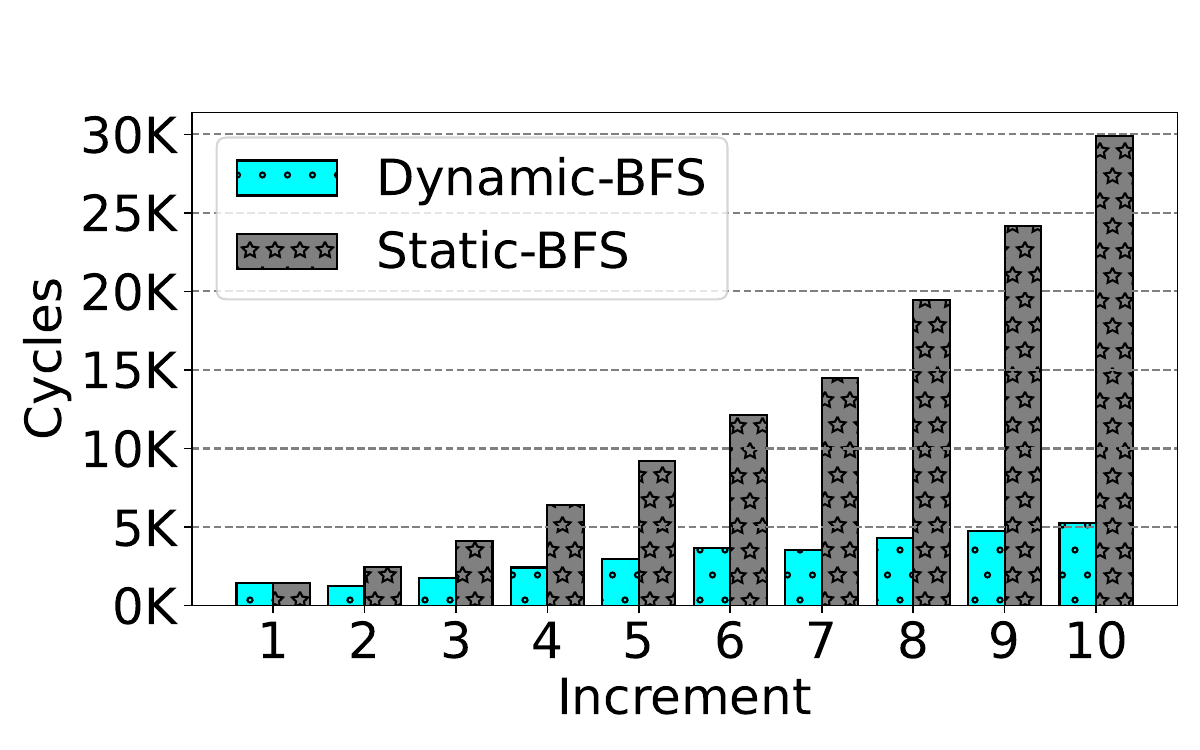}
    \caption{Snowball Sampling \#V: 50K.}
    \label{fig:snowball-sample-50k}
  \end{subfigure}
  \hfill
  \begin{subfigure}{0.49\linewidth}
    \centering
    \includegraphics[width=\linewidth]{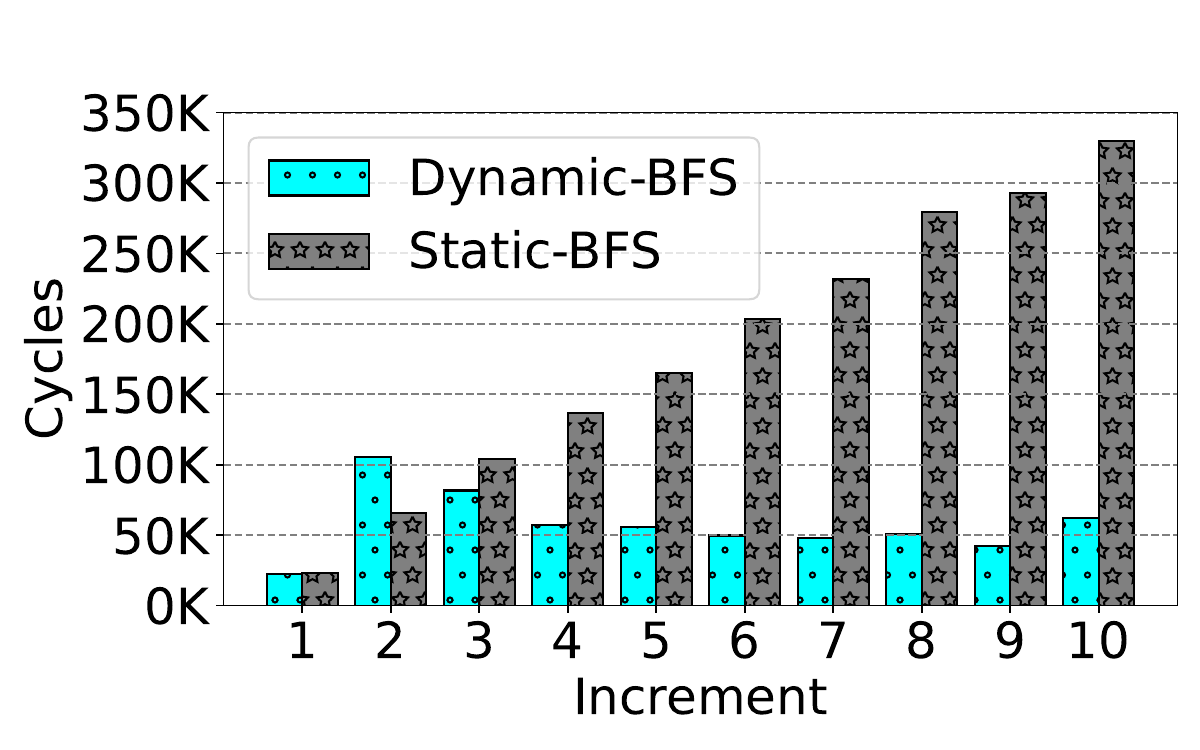}
    \caption{Edge Sampling \#V: 500K.}
    \label{fig:edgesample-500K-32}
  \end{subfigure}
  \hfill
  \begin{subfigure}{0.49\linewidth}
    \centering
    \includegraphics[width=\linewidth]{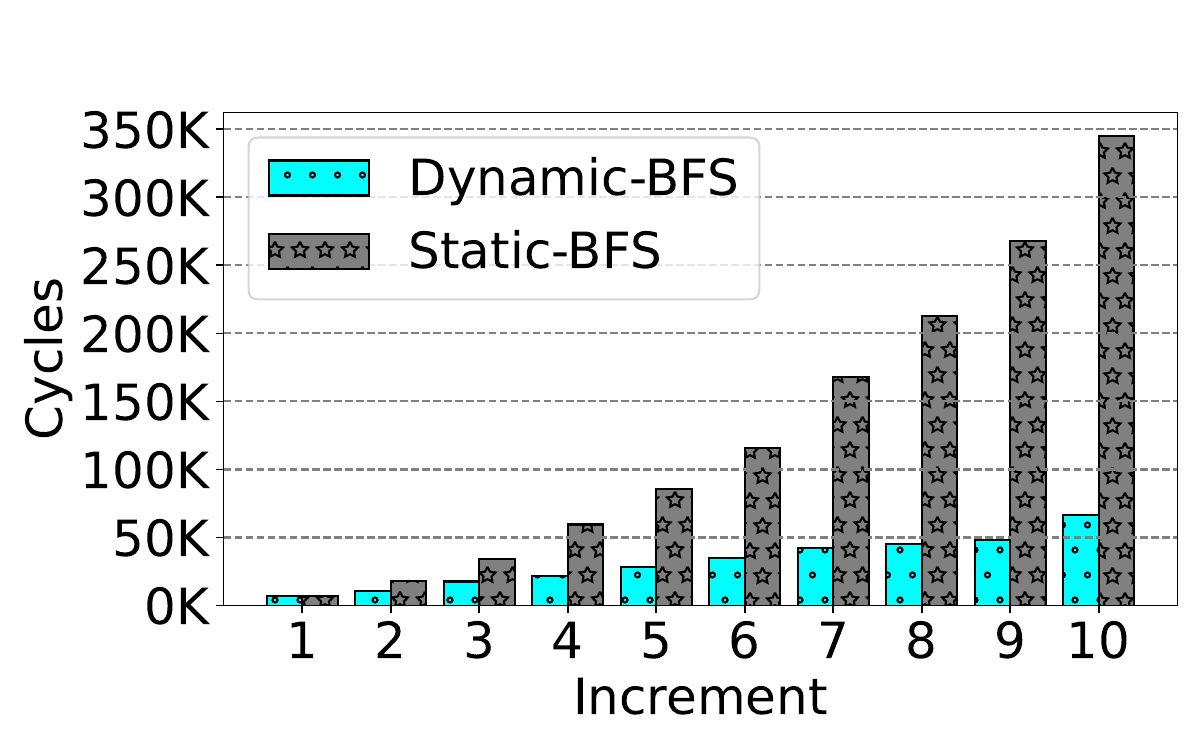}
    \caption{Snowball Sampling \#V: 500K.}
    \label{fig:snowball-500K-32}
  \end{subfigure}
  \caption{Time taken in simulation cycles on a $32 \times 32$ chip.}
  \label{fig:dynamic-bfs-performance}
\end{figure}

\begin{figure}
  \centering
  \begin{subfigure}{0.49\linewidth}
    \centering
    \includegraphics[width=\linewidth]{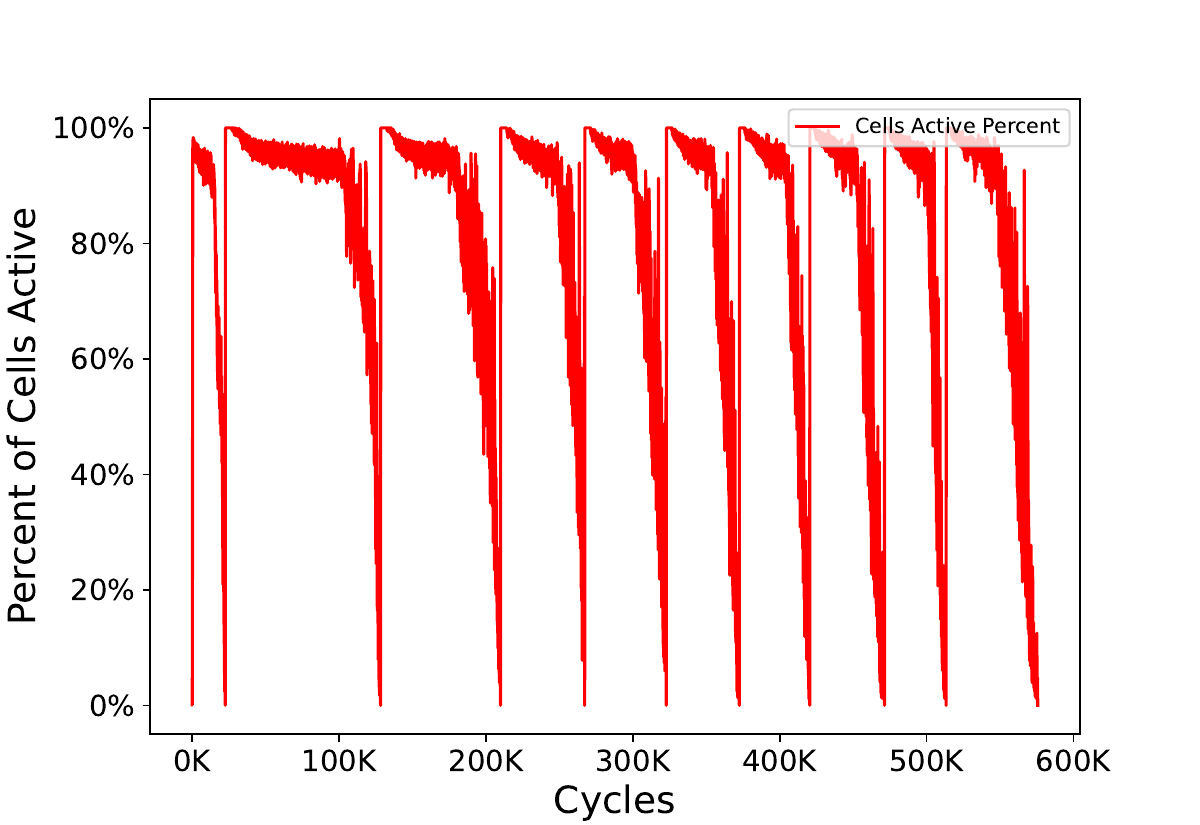}
    \caption{Edge Sampling 500K.}
    \label{fig:dynamic-active-edge-500K}
  \end{subfigure}
  \hfill
  \begin{subfigure}{0.49\linewidth}
    \centering
    \includegraphics[width=\linewidth]{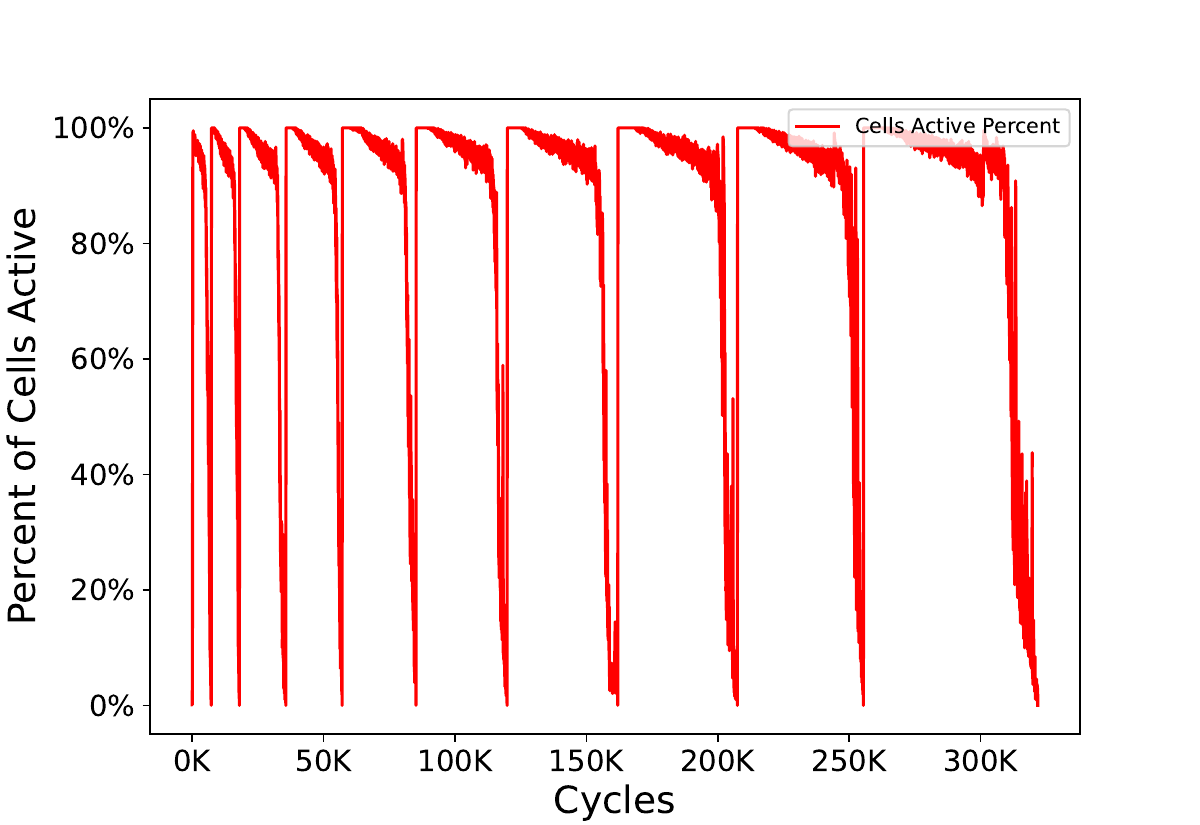}
    \caption{Snowball Sampling 500K.}
    \label{fig:dynamic-active-snowball-500K}
  \end{subfigure}
  \caption{Activation status of compute cells for Dynamic-BFS execution on a $32 \times 32$ chip.}
  \label{fig:dynamic-active}
\end{figure}

Figure \ref{fig:dynamic-bfs-performance} shows the performance of Dynamic-BFS, which uses results from prior executions. It is compared against a reference called Static-BFS, providing context on the execution time it would have taken had it run from scratch. Figure \ref{fig:dynamic-active} shows the activation status of the CCs as the execution progresses. It shows that there is a complete pause after a batched increment is processed and before a new batched increment begins, highlighting the nature of batched processing. Dynamic-BFS consumed less time on graphs constructed with Snowball Sampling compared to Edge Sampling. This is due to the nature of Snowball Sampling, where new edges are inserted in a BFS manner as opposed to random edge sampling. This results in many of the paths taken by BFS being the true paths, thus creating fewer \textit{actions}. Compared to Snowball Sampling, Dynamic-BFS with Edge Sampling created $1.47\times$ and $1.31\times$ more \textit{actions} for graph sizes of $50K$ and $500K$, respectively.

Since the experiments pertain to batched updates, the data transfer time is not included as it is the same for both. However, data transfer will have an impact under streaming dynamic updates. In that case when an edge is inserted, it will instantly spawn a new \textit{action}, thus creating more work at the compute and network of the CCs, thereby interfering with data transfer. This complex interaction will be addressed and examined in future studies.

\section{Conclusion \& Future Work}

The work discuss in this paper aims to provide a unified approach to system design for irregular memory applications such as graph processing. It achieves this by providing architectural support for massive fine-grain message-driven computations that send \textit{actions} carrying ``work to data''. This underlying raw parallelism is transparently exported to the application by the means of an asynchronous programming model and a runtime system that manage an inherently parallel and dynamic object model as the data structure of the irregular input graph. 

We identify the following areas of research direction as a natural progression from the work discussed here:

\textbf{Reducing Average Latency of the NoC:} Techniques to reduce the diameter of the mesh network, such as augmenting the mesh with extra channels per CC, like Ruche Networks \cite{Ruche2020}, must be explored. These extra channels directly connect CCs to distant CCs, thereby lowering the mesh network diameter and potentially reducing congestion.

\textbf{Adaptive Routing Algorithms:} In the current experimental setup, we employed turn-restricting dimension-ordered routing, which guarantees deadlock freedom but cannot address or mitigate congestion, even when less congested paths exist in the mesh network. Adaptive routing algorithms may address and mitigate congestion more effectively and together may also aid in exploring wafer-scale designs of the system.

\textbf{Wafer-Scale Designs:} The system is composed of homogeneous compute cells connected by a mesh network that has a higher degree of tolerance for manufacturing and runtime faults. Adaptive routing, combined with lower diameter mesh techniques such as the Ruche network, has the potential to enable wafer-scale deployments with high memory capacity for processing larger graphs.

%
%
%
\bibliographystyle{splncs04}
\bibliography{Reference}
%

\end{document}